\documentclass[manuscript,screen]{acmart}
\AtBeginDocument{%
  }

\usepackage{tcolorbox}
\usepackage{alltt}
\usepackage{subcaption}
\usepackage{tabularx}
\usepackage{booktabs}
\usepackage{multirow}

\setcopyright{acmlicensed}
\copyrightyear{2026}
\acmYear{2026}
\acmDOI{XXXXXXX.XXXXXXX}
\begin{document}



\title[From Incomplete Architecture to Quantified Risk: Multimodal LLM-Driven Security Assessment for CPSs]{From Incomplete Architecture to Quantified Risk: Multimodal LLM-Driven Security Assessment for Cyber-Physical Systems}


\author{Shaofei Huang}
\email{sf.huang.2023@smu.edu.sg}
\orcid{0009-0002-3954-1843}
\affiliation{%
  \institution{Singapore Management University}
  \city{Singapore}
  \country{Singapore}
}

\author{Christopher M. Poskitt}
\email{cposkitt@smu.edu.sg}
\orcid{0000-0002-9376-2471}
\affiliation{%
  \institution{Singapore Management University}
  \city{Singapore}
  \country{Singapore}
}

\author{Lwin Khin Shar}
\email{lkshar@smu.edu.sg}
\orcid{0000-0001-5130-0407}
\affiliation{%
  \institution{Singapore Management University}
  \city{Singapore}
  \country{Singapore}
}


\renewcommand{\shortauthors}{Huang et al.}

\begin{abstract}

Cyber-physical systems often contend with incomplete architectural documentation or outdated information resulting from legacy technologies, knowledge management gaps, and the complexity of integrating diverse subsystems over extended operational lifecycles. This architectural incompleteness impedes reliable security assessment, as inaccurate or missing architectural knowledge limits the identification of system dependencies, attack surfaces, and risk propagation pathways. To address this foundational challenge, this paper introduces ASTRAL (Architecture-Centric Security Threat Risk Assessment using LLMs), an architecture-centric security assessment technique implemented in a prototype tool powered by multimodal LLMs. The proposed approach assists practitioners in reconstructing and analysing CPS architectures when documentation is fragmented or absent. By leveraging prompt chaining, few-shot learning, and architectural reasoning, ASTRAL extracts and synthesises system representations from disparate data sources. By integrating LLM reasoning with architectural modelling, our approach supports adaptive threat identification and quantitative risk estimation for cyber-physical systems. We evaluated the approach through an ablation study across multiple CPS case studies and an expert evaluation involving 14 experienced cybersecurity practitioners. Practitioner feedback suggests that ASTRAL is useful and reliable for supporting architecture-centric security assessment. Overall, the results indicate that the approach can support more informed cyber risk management decisions.

\end{abstract}




\maketitle

\section{Introduction}
\label{sec:Introduction}

In cyber-physical systems (CPSs), software architecture forms the structural foundation upon which system functionality, performance, and security depend. Architectural models play a pivotal role in security assessment by providing a coherent representation of components, interfaces, and data flows within complex, multi-layered environments~\cite{Jiang2024}. Decisions made at the architectural level, such as component decomposition, communication pathways, and trust boundary definition, directly shape the system's attack surface and influence how vulnerabilities emerge and propagate. These structural choices determine not only where defensive controls can be placed, but also how effectively threats can be isolated, contained, or mitigated. Architectural patterns that emphasise isolation, redundancy, and secure communication therefore enhance system resilience and reduce the likelihood of cascading failures. Effective security assessment is thus inseparable from architectural analysis, as the same structural decisions that govern system behaviour and performance also define the conditions under which security controls can be systematically integrated and enforced throughout the system life-cycle.

In real-world settings, practitioners frequently encounter significant impediments due to \textit{architectural incompleteness}, where documentation is fragmented, outdated, or entirely absent~\cite{Hofer2018architecture}. This gap in architectural knowledge disrupts security assessment workflows, particularly within frameworks that presuppose the existence of exhaustive architectural documentation. CPS environments often contend with these gaps due to extended operational life-cycles, persistence of legacy technologies, and complexities of multi-vendor subsystem integration. These factors result in missing details about components, versions, and interconnections \cite{Yang2023,Wolf2015}. Furthermore, threat models developed during the initial design phase rapidly become obsolete as the physical and digital architectures evolve through iterative updates \cite{Huang2024a, Jamil2021automated}, creating a discontinuity that complicates operational security assessment. Such architectural gaps hinder effective remediation, as ad hoc mitigations may inadvertently cause further system degradation \cite{Bakirtzis2020}. Recent advancements in Large Language Models (LLMs) \cite{Zhang2024} suggest new possibilities for reasoning over and reconciling architectural gaps. By leveraging the ability to process multimodal inputs, such as architecture and network topology diagrams \cite{Xu2024a}, LLMs exhibit the potential to reconstruct architectural contexts from fragmented data. A coherent architectural representation, if reliably established, could in turn provide the structural foundation for downstream security workflows, including continuous threat modelling (e.g., STRIDE-LM \cite{Muckin2014}), dynamic attack tree synthesis \cite{Schneier1999}, and adaptive risk assessment across the CPS life-cycle. However, the systematic integration of these capabilities into a unified, architecture-centric CPS security assessment workflow remains underexplored.

\begin{figure}[t]
\centering
\includegraphics[width=0.9\textwidth]{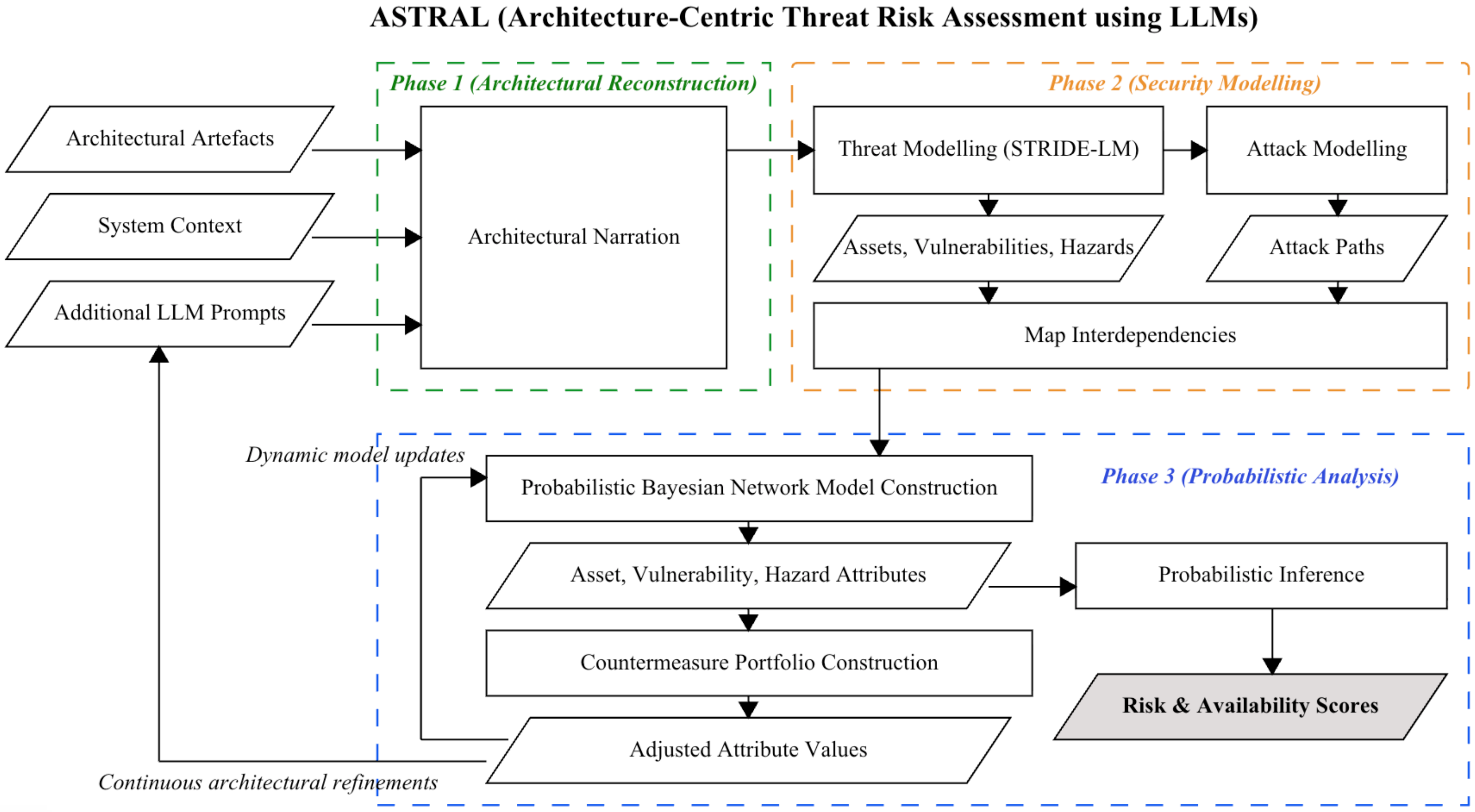}
\Description[Workflow diagram of ASTRAL framework]{CPS security assessment workflow using LLMss.}
\caption{Overview of ASTRAL. The security assessment workflow for CPSs with incomplete architectural knowledge, enabled by multimodal LLMs, progresses through three phases: architectural reconstruction (Phase 1); security modelling (Phase 2); and probabilistic analysis (Phase 3).}
\label{fig:ASTRAL_Overview}
\end{figure}

ASTRAL (\underline{A}rchitecture‑Centric \underline{S}ecurity \underline{T}hreat \underline{R}isk \underline{A}ssessment using \underline{L}LMs), a multimodal LLM‑driven technique for architecture‑centric security assessment of CPSs, operationalises these capabilities via the multi-phase workflow illustrated in Figure~\ref{fig:ASTRAL_Overview}. ASTRAL supports practitioners in reconstructing and subsequently analysing these architectures when documentation is incomplete or absent. In this context, \textit{reconstruction} refers to the systematic derivation of a coherent architectural representation from fragmented and heterogeneous artefacts. The methodology integrates prompt chaining, few‑shot learning, and structured LLM reasoning \cite{Sahoo2024} to systematically extract and organise architectural knowledge for security assessment. By combining LLM reasoning with architecture-centric modelling, our research 
facilitates the dynamic identification of threats alongside quantitative risk assessment within a unified, life-cycle-integrated framework.

\subsection{Research Questions}

To evaluate the effectiveness and practical applicability of our proposed approach, we investigate the following research questions:
\begin{itemize}
    \item \textbf{RQ1:} \textit{To what extent do ASTRAL’s key design elements contribute to architectural reconstruction quality and structural validity?} This research question evaluates how ASTRAL’s key design elements (multimodality, guardrails, sample configuration settings) influence the quality and structural validity of reconstructed architectures. We conduct ablation experiments to quantify the relative degradation in trust boundary and attack path validity, graph connectivity, and semantic coherence when each element is selectively disabled or modified.
    
    \item \textbf{RQ2:} \textit{What architectural information gaps can ASTRAL fill?} This research question examines the extent to which ASTRAL can address incomplete architectural information using both textual data and images. The findings will demonstrate how LLM‑assisted reconstruction of architectural knowledge can help bridge the information gaps between threat modelling, attack modelling, and risk analysis stages within the security assessment workflow.
    
    \item \textbf{RQ3:} \textit{How do cybersecurity practitioners perceive the trustworthiness and reliability of the security assessment results generated by ASTRAL?} This research question explores the degree to which experienced cybersecurity practitioners perceive LLM‑generated results as trustworthy and reliable. The findings will provide insight into the credibility, consistency, and correctness of LLM‑assisted security assessments in CPS contexts.
    
    \item \textbf{RQ4:} \textit{How useful is our proposed approach for CPS practitioners?} This research question assesses the extent to which practitioners view LLM‑assisted security assessment approaches as useful to their work. The results will help determine the practical applicability, perceived benefits, and adoption potential of ASTRAL for real‑world CPS security assessments.

\end{itemize}

\begin{figure*}[ht]
\centering
\begin{subfigure}{0.5\textwidth}
\centering
\includegraphics[width=\textwidth]{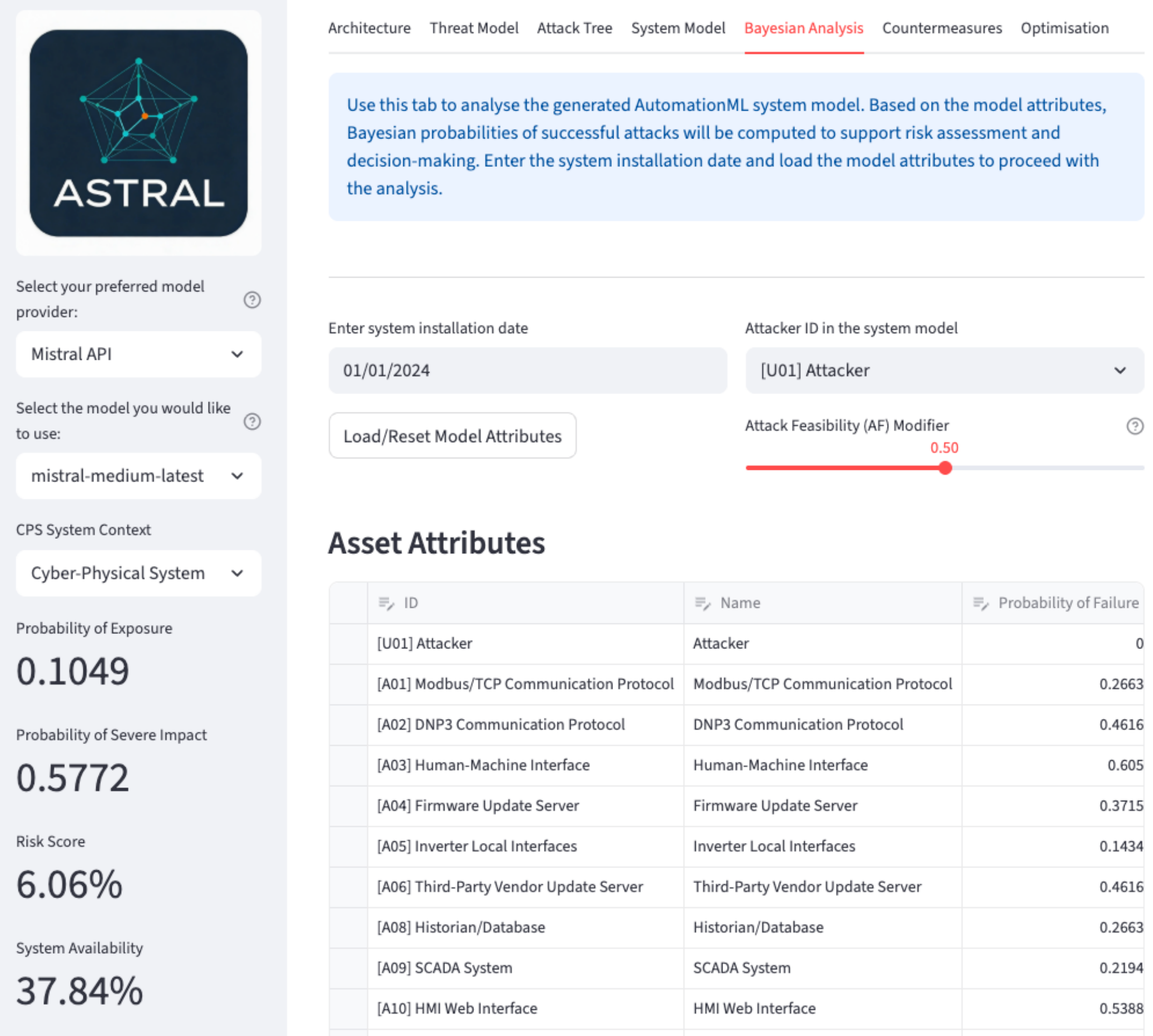}
\end{subfigure}
\begin{subfigure}{0.41\textwidth}
\centering
\includegraphics[width=\textwidth]{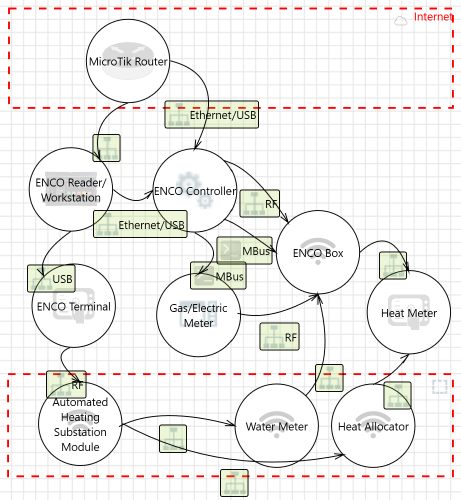}
\end{subfigure}
\caption{Prototype implementation of ASTRAL (left) with architectural information reconstructed from incomplete artefacts, e.g., data flow diagrams (DFDs) (right). Source code and demo available at~\cite{ASTRAL2026}.}
\label{fig:ASTRAL}
\end{figure*}

\subsection{Contributions}

The main contributions of our work are:

\begin{itemize}

    \item \textbf{Addressing architectural incompleteness as a core CPS challenge}: We advance the understanding of how incomplete architectural knowledge impacts CPS security assessment, and propose applied methods that integrate information processing, system architecture, and risk analysis to address security challenges common across CPSs. We integrate threat and attack models directly into a Bayesian Network (BN) framework to evaluate critical CPS assets, vulnerabilities, and hazards within specific architectural contexts. This integration enables rigorous, quantitative, and scenario-driven risk analysis tailored for the uncertainties of CPS environments~\cite{Huang2026}.
    
    \item \textbf{Introducing a novel security assessment approach}: We introduce ASTRAL, a novel architecture‑centric security assessment approach that employs multimodal LLMs to automatically reconstruct and synthesise system architecture from outdated or incomplete documentation. This ensures the capture and availability of architectural knowledge essential to link all stages of assessment within complex CPS environments. By leveraging structured LLM augmentation and practitioner-validated tooling for the reconstruction and interpretation of architectural knowledge, we bridge information gaps between threat modelling, attack modelling, and risk analysis stages in the security assessment workflow. This represents a more comprehensive methodology compared to prior studies that have applied LLMs only to isolated aspects of threat modelling~\cite{Adams2024, Sidhant2024}.
    
    \item \textbf{Developing a practitioner-oriented prototype tool}: We develop a prototype tool (Figure~\ref{fig:ASTRAL}) that enables adaptive, real‑time security assessment and decision support for CPSs with incomplete architectural knowledge. Designed to support practical needs of CPS practitioners, the tool demonstrates the feasibility and practical potential of implementing the ASTRAL approach within real-world CPS environments.
    
    \item \textbf{Empirical validation with practitioners}:  We conduct systematic validation with 14 security practitioners, including Chief Information Security Officers (CISOs), to evaluate ASTRAL's usability, trustworthiness, and reliability in assessing security risks and prioritising mitigation strategies within realistic CPS scenarios. Their feedback provided qualitative insights that complement and extend existing LLM-based approaches to threat and risk assessment in CPS contexts~\cite{Salek2025, Gultekin2025}.
    
\end{itemize}

The remainder of the paper is structured as follows. Section~\ref{sec:Background} provides background and discusses related work on CPS security modelling frameworks and LLMs. Section~\ref{sec:Methodology} details our methodology for LLM-enabled security assessment, illustrated with a running example. Section~\ref{sec:case_studies} presents evaluations across other CPSs that feature contrasting architectural characteristics and documentation gaps. Section~\ref{sec:Ablation_Study} presents an ablation study validating the necessity of ASTRAL’s key design elements through controlled degradation experiments on architectural artefacts. Section~\ref{sec:Validation_From_Practitioners} describes the user study with cybersecurity practitioners and presents the analysis of survey results. Section~\ref{sec:Conclusion} concludes with a summary of key contributions from our research.

\section{Background and Related Work}
\label{sec:Background}

In this section, we present the foundational concepts that underpin our study and discuss related work.

\subsection{CPS Security Modelling Frameworks}

CPSs are engineered systems that integrate computational elements with physical processes. They utilise embedded sensors, actuators, and communication networks for real-time monitoring and control, enabling automated and intelligent interaction between digital and physical environments. CPSs underpin critical infrastructure sectors such as energy, transportation, healthcare, and manufacturing, where safety, reliability, and security are essential. The complexity of CPSs stems from integration of diverse hardware, software, and network components operating under stringent timing and safety constraints. Addressing these challenges requires specialised architectural modelling and security assessment methodologies tailored to CPS functionality and cybersecurity needs.

Security modelling is an important process that examines a system’s architecture or design to identify and mitigate vulnerabilities \cite{Huang2024a}. In CPSs, this involves systematically identifying and evaluating vulnerabilities and threats that may impact confidentiality, integrity, availability, and domain-specific properties such as safety and reliability. Building on Microsoft’s original STRIDE methodology \cite{Hernan2005} -- encompassing Spoofing, Tampering, Repudiation, Information Disclosure, Denial of Service, and Elevation of Privilege -- the STRIDE-LM framework \cite{Muckin2014} adds Lateral Movement, further supporting structured identification and mitigation of threats in complex CPS architectures. Attack trees provide a complementary approach for modelling adversarial behaviour, offering a hierarchical representation that decomposes attacker objectives into possible attack paths and sub-goals \cite{Schneier1999}.

However, these security models often lack the ability to represent the dynamic, multi-stage nature of CPS attacks~\cite{Huang2024a}. Existing risk quantification approaches frequently overlook uncertainties that arise from incomplete data, changing system states, or sensor inaccuracies, all of which are common in CPS environments~\cite{Makelburg2025uncertainty}. Moreover, architectural documentation is frequently incomplete or obsolete, a consequence of the extended service life of these systems~\cite{Cardenas2019}. This lack of structural fidelity significantly degrades the accuracy of security assessments, particularly as systems evolve and novel threat vectors emerge.

To address these uncertainties, probabilistic models such as BNs offer a powerful framework for real-time inference and dynamic risk quantification \cite{Kalantarnia2009, Poolsappasit2011}. By integrating diverse heterogeneous data sources, BNs enable adaptive security evaluations that reflect the changing system conditions and threat landscape \cite{Huang2026}. For semantic interoperability across diverse CPS implementations, BN architectures can be encoded using domain-specific languages such as Automation Markup Language (AutomationML), a standardised XML-based format used for the exchange of engineering data in industrial automation systems~\cite{AutomationML2025}.

While existing frameworks such as STRIDE-LM and attack trees excel at structured threat enumeration, they typically presuppose exhaustive architectural knowledge. This assumption is often untenable for long-lived CPSs characterised by outdated or incomplete documentation~\cite{Jamil2021automated}. Similarly, although BNs effectively manage probabilistic risk, they require pre-existing topologies to function~\cite{Zebrowski2022bayesian}. ASTRAL distinguishes itself by prepending a multimodal LLM reconstruction phase to generate these structural inputs from partial artefacts. This ensures a level of workflow continuity that remains absent in traditional, static assessment approaches.

\subsection{Large Language Models}

LLMs are Transformer-based models~\cite{Vaswani2017} trained on large-scale corpora to learn rich semantic and syntactic representations of language~\cite{Zhao2023}. Through multi-layer attention mechanisms, they capture long-range dependencies and contextual relationships, enabling complex reasoning across heterogeneous inputs. Beyond standard NLP tasks, LLMs can be adapted via prompt engineering, few-shot learning, and fine-tuning to perform structured analysis and decision support. Retrieval-Augmented Generation (RAG)~\cite{Lewis2020} further extends these capabilities by incorporating external knowledge sources, improving grounding and contextual relevance.

Recent advancements have produced multimodal LLMs capable of jointly reasoning over textual and visual inputs, enabling the integration of heterogeneous artefacts into coherent semantic representations. In cybersecurity and software engineering, multimodal LLMs show significant promise for the automated extraction of structured knowledge from unstructured artefacts~\cite{Xu2024a}, dynamic threat and risk modelling~\cite{Elsharef2024,Gultekin2025,Salek2025}, and scalable decision support that synthesises both threat intelligence and system state information~\cite{Tian2025}.


Recent work underscores the growing role of Large Language Models (LLMs) in cybersecurity and their integration within cyber-physical systems (CPS). Xu et al. \cite{Xu2024a} survey LLM-based cybersecurity applications, highlighting a shift toward autonomous agents for multi-stage security workflows, while Xu et al. \cite{Xu2024b} propose a taxonomy of LLM-enabled CPS research by application domain and LLM function. Beyond these surveys, LLMs support diverse security workflows, including LLM-driven threat modelling \cite{Elsharef2024}, architecture-based penetration test generation \cite{Sarvejahani2025}, and RAG-enhanced CPS risk assessment in autonomous forestry machinery \cite{Gultekin2025}. Additional examples include automated threat modelling for banking systems \cite{Wu2024} and MITRE ATT\&CK-based LLM frameworks for transportation CPSs \cite{Salek2025}.

Despite these significant developments, a critical methodological gap remains: existing LLM-based approaches generally presuppose the availability of comprehensive architectural data. This reliance fails to address architectural incompleteness, which serves as a fundamental barrier to downstream security analysis. ASTRAL departs from these works by systematically sequencing multimodal reconstruction with threat and attack modelling, followed by risk quantification. By leveraging heterogeneous inputs and prompt chaining, ASTRAL facilitates a life-cycle-integrated CPS assessment that remains robust even when architectural documentation is incomplete or absent. Our work enables the use of multimodal architectural artefacts -- including diagrams and unstructured documentation -- to support security assessment across diverse CPS deployments, such as Automotive In-Vehicle Infotainment (IVI) systems~\cite{Huang2024b}, solar power systems~\cite{Forescout2025}, and municipal heating systems~\cite{DragosFrostyGoop2024, Unit42FrostyGoop2024}.

\section{Methodology of ASTRAL}
\label{sec:Methodology}

As shown in Figure~\ref{fig:ASTRAL_Overview}, the ASTRAL workflow begins with architectural reconstruction, leveraging architectural artefacts to construct an \emph{architectural narration} -- a structured textual description that systematically identifies and analyses artefact elements to support threat and attack modelling. These elements include attacker or attack-capable entities; key components such as PLCs (Programmable Logic Controllers), sensors, and actuators; trust boundaries and operational zones; data flows and interactions; technologies and protocols; critical cyber-physical assets and functions; and attack entry points. This narration is then encoded into LLM prompts for security modelling, which generates both a threat model and an attack model. These outputs are subsequently synthesised to build probabilistic models encoded in AutomationML format, facilitating quantitative risk analysis by enabling probabilistic inference of risk factors and outcomes.

To illustrate this workflow, we utilise the \textit{FrostyGoop} attack scenario \cite{DragosFrostyGoop2024, Unit42FrostyGoop2024} as a running example. Discovered in 2024, \textit{FrostyGoop} represents the first documented instance of OT-centric malware using Modbus TCP communications to directly disrupt municipal heating services, resulting in a two-day outage for over 600 apartment buildings in Ukraine. This incident highlights the critical challenge of architectural incompleteness, as public documentation for such systems is often fragmented or entirely absent. In the following subsections, we demonstrate how ASTRAL transforms these limited, security-relevant artefacts into a structured architectural narration that drives threat analysis, attack path analysis, and BN-based risk estimation.

A video demonstration of the running example, showcasing the analysis of FrostyGoop using the ASTRAL tool, is available at the following YouTube \href{https://www.youtube.com/watch?v=T_5pfyUnrJc}{link}.

\subsection{Preliminaries}

\subsubsection{Data Sources}

ASTRAL is designed to support architecture‑centric security assessment across a wide range of artefact types. These may include image‑based representations of architectural diagrams (e.g., PNG, JPG, BMP, GIF), configuration files, asset inventories, and potentially live operational data. While the framework is likely extensible to all such sources, the current prototype implements its core functionality using only image‑based artefacts.

In this study, the primary artefacts comprise data flow diagrams (DFDs) generated using tools such as the Microsoft Threat Modelling Tool \cite{MicrosoftTMT2024}. Supplementary schematic, network, and architectural diagrams are also incorporated, extracted from security advisories, incident reports, and technical documentation. Reconstruction is performed using the provided image‑based artefacts, augmented by practitioner‑supplied system context. Consequently, the fidelity of both the reconstructed architecture and subsequent risk estimation depends on the quality, granularity, and completeness of the supplied artefacts.



\emph{\textbf{Running Example.}} In the case of the \textit{FrostyGoop} attack scenario, the available documentation exemplifies the challenge of architectural incompleteness. Open‑source threat reports and advisories provide only a coarse description of the municipal heating architecture, including logical zones, supervisory components, field devices and remote terminal units, and the use of Modbus TCP between the control and field layers \cite{DragosFrostyGoop2024}. Important architectural details, including device vendor and firmware versions, internal subnet structures, firewall rule sets, and the exact placement of remote access gateways, are either missing or described only informally. To address this documentation gap, ASTRAL ingests an incomplete DFD (as depicted in Figure \ref{fig:ASTRAL}), constructed using the Microsoft Threat Modelling Tool based on information from available open-source intelligence. It then systematically reconstructs an architectural narration from the image-based artefact. Through this process, ASTRAL explicitly enumerates CPS assets, trust boundaries, and security‑relevant data flows, thereby transforming an incomplete architectural artefact into a structured model suitable for subsequent security analysis.

\subsubsection{LLM}

ASTRAL is designed to be LLM provider- and model-agnostic, supporting Mistral AI, Google Gemini, Anthropic Claude, and OpenAI models. It leverages LangChain's consistent API framework to enable seamless switching of models and providers across workflow phases. This abstraction layer addresses API syntax differences between LLM providers, allowing practitioners to adapt prompts and chaining logic while maintaining methodological portability across CPS scenarios. For this research, we primarily selected Mistral AI's \textit{Mistral Medium 3.1}~\cite{MISTRAL_Medium}. This model excels at processing multimodal inputs and handling complex security modelling tasks -- capabilities that are essential for CPS architectural reconstruction and timely security assessment.

\subsection{Security Assessment Workflow}

Structured prompt engineering is a central aspect of our methodology, systematically guiding the LLM through successive phases of the security assessment workflow, comprising: architectural reconstruction, security modelling, and probabilistic analysis~(Figure~\ref{fig:ASTRAL_Overview}). These three integrated phases are described in the next sections.

\subsubsection{Phase 1: Architectural Reconstruction}

ASTRAL employs multimodal inputs to construct a coherent architectural narration from incomplete architecture, grounded in a specified system context and LLM reasoning guardrails. To guide reasoning, we leverage domain-specific ontologies to extract contextual evidence from available CPS documentation. The model integrates heterogeneous data sources, such as fragmented diagrams and unstructured text, to fill knowledge gaps through pattern completion and constrained inference.

Specifically, pattern completion matches partial architectural elements against established CPS reference architectures, such as Purdue Enterprise Reference Architecture~\cite{Williams1994} -- to infer missing components via structural analogy. Constrained inference applies semantic guardrails and ontological rules (e.g., IEC 62443 zone mappings~\cite{IEC62443} and AutomationML role classes~\cite{AutomationML2025}) to limit LLM generation to plausible configurations -- architectural arrangements that are structurally and semantically valid within CPS standards. Plausible configurations include Purdue Level 2 zones containing only PLCs, HMIs (Human-Machine Interfaces), and historians (never enterprise servers), OPC UA (OPC Unified Architecture) servers residing exclusively in demilitarised zones between Purdue Levels 3 and 2, or field devices (Level 0/1) communicating solely via industrial protocols (e.g., Modbus, PROFINET) rather than HTTP. This approach ensures the reconstruction remains consistent with established CPS architectural patterns rather than relying on unguided text generation.

During the initial architectural reconstruction in Phase 1, the LLM extracts and synthesises information from uploaded architectural artefacts using an initial prompt. The process is further strengthened through iterative prompting by security practitioners, enabling interactive refinement of LLM outputs. Practitioners may, for instance, instruct the model to “closely examine all text labels in the diagram and identify product names and versions.” Such iterative engagement enables the LLM to reconstruct precise textual representations of the system architecture, thereby supporting robust threat and attack modelling in the subsequent stages even in cases of architectural incompleteness. 

To ensure schema compliance and semantic consistency during model generation, we employ LLM \textit{prompt chaining}~\cite{Sun2024prompt}, a prompt engineering technique where the output from one LLM prompt (or step) is used as the input for the next, creating a chain that leads to a more accurate and high-quality final result. This process aligns model outputs with the IEC 62714 standard~\cite{IEC62714-1-2018}, which defines the AutomationML format, and its corresponding security extensions~\cite{Eckhart2020, AutomationMLCPS2025}. These extensions provide the necessary security semantics for the systematic representation and automated assessment of CPS security risks.

The prompt chaining approach combines one- or few-shot examples with semantic guardrails and controlled sampling configurations. Semantic guardrails define rules and contextual boundaries that direct the model to produce schema-aligned and logically consistent outputs, while controlled sampling configurations adjust the \textit{temperature} to control randomness (a lower value yields more predictable output while a higher value allows for more creativity), and modify \textit{top\_p} to control the model output by augmenting the vocabulary size (values close to 0 restrict output to the most probable tokens, making it more deterministic, whereas values close to 1 allow a larger portion of the vocabulary to be considered, making the output more random). Together, these mechanisms mitigate hallucination risks~\cite{Huang2025hallucination}, enhance output consistency, and enable the LLM to handle incomplete or ambiguous architectural data more effectively.

\begin{figure}[ht]
\centering
\begin{minipage}{0.9\textwidth}
\begin{alltt}\footnotesize
You are a Senior Solution Architect tasked with narrating a system architectural diagram (e.g., Data
Flow Diagram) to a Senior Security Architect experienced in IEC 62443 and the Purdue model. Your narration supports threat \\modelling and attack tree development for a cyber-physical system, even if the architecture appears IT-centric.

System context: Municipal heating system

Controlled sampling configuration:
- temperature = 0.1
- top_p = 0.9

Think deeply to thoroughly analyse the diagram and provide a structured narration strictly based on visible content, \\covering:
1. Attacker or Attack-Capable Entities (explicit or implied, e.g., adversaries, operators)
2. Key Components (systems, devices, applications, network infrastructure, sensors, actuators, OT assets)
3. Trust Boundaries and Purdue Zones
4. Data Flows and Interactions (including protocols, data types, communication links)
5. Technologies, Platforms, and Standards
6. Assets and Functions with cyber-physical significance (PLCs, controllers, field devices, routers, meters, etc.)
7. Attack Entry Points (explicit or implied entities that could initiate attacks)
8. Any other architectural details supporting threat modelling and attack tree development

Structure your response using these exact section headers only:
- Attacker or Attack-Capable Entities
- Key Components
- Trust Boundaries and Purdue Zones  
- Data Flows \& Interactions  
- Technologies and Protocols  
- Assets and Functions  
- Attack Entry Points  

IMPORTANT - Follow these strictly enforced semantic guardrails:
- Base your narration solely on the provided diagram; do not infer or assume details beyond what is visible.
- Do not start or end with commentary or extra text.
- Do not infer or guess beyond what is visibly present.
- Do not provide recommendations—only factual narration.
- Use only the specified headers and no additional formatting.
\end{alltt}
\end{minipage}
\caption{Architectural reconstruction prompt with semantic guardrails and controlled sampling. Generates structured architectural narration by extracting attackers, CPS components, Purdue trust zones, data flows, protocols, assets, and attack entry points from architectural artefacts to enable automated IEC 62443 threat modelling and attack tree generation.}
\label{fig:LLM-Prompt-Arch}
\end{figure}

\emph{\textbf{Running Example.}} ASTRAL first reconstructs the CPS architecture from incomplete artefacts in the FrostyGoop scenario, using an initial LLM prompt to reconstruct the architectural narration, as shown in Figure~\ref{fig:LLM-Prompt-Arch}. A controlled sampling configuration, with a \textit{temperature} of 0.1 (promoting deterministic, schema-consistent generation) and a \textit{top\_p} value of 0.9 (allowing a larger portion of the vocabulary to be considered), is specified in the prompt to mitigate hallucination risks. The prompt specifies the persona of a senior solution architect instead of a cybersecurity expert, as the objective of ASTRAL’s architectural reconstruction phase is to generate an accurate and comprehensive structural narration of the CPS rather than a threat-oriented assessment. This perspective encourages the model to emphasise system composition, inter-component dependencies, and integration boundaries, which are essential for subsequent security analysis. The use of an architecture-focused persona thus ensures that the generated narration serves as a reliable foundation for downstream risk reasoning.

Semantic guardrails and system context are also specified. For example, the prompt includes specific instructions for the model to “\textit{think deeply to thoroughly analyse the diagram and provide a structured narration strictly based on visible content}” and to “\textit{base [its] narration solely on the provided diagram; do not infer or assume details beyond what is visible.}” The prompt further constrains the output to “\textit{use only the specified headers and no additional formatting}” and to structure the response under headings such as “\textit{Attacker or Attack-Capable Entities}”, “\textit{Trust Boundaries and Purdue Zones}”, and “\textit{Data Flows \& Interactions}”, thereby guiding the LLM towards accurate and schema-consistent architectural reconstruction.

\subsubsection{Phase 2: Security Modelling}

Building upon the output from Phase 1, ASTRAL uses tailored prompts to identify threats across the seven STRIDE-LM categories: Spoofing, Tampering, Repudiation, Information Disclosure, Denial of Service, Elevation of Privilege, and Lateral Movement~\cite{Hernan2005,Muckin2014}. The LLM synthesises architectural and behavioural data relevant to CPS environments, enabling threat elicitation. Building on the analysis, the LLM generates attack trees and attack paths, capturing causal sequences, attacker objectives, and exploitation methods within the system architecture. This process enables dynamic representation of attacker behaviour and potential attack flows.

\emph{\textbf{Running Example.}} In this security modelling phase, ASTRAL utilises LLM prompts to generate the threat model and attack tree, respectively. These prompts are shown in Figures~\ref{fig:LLM-Prompt-ThreatModel} and \ref{fig:LLM-Prompt-AttackTree}. Within the context of threat and attack modelling, hallucination risks inherent in the LLM are a key concern, as the fidelity and accuracy of the generated outputs have a substantial influence on subsequent analytical outcomes. To mitigate these risks, a controlled sampling configuration is employed, using a \textit{temperature} of 0.1 (ensuring deterministic and precise outputs, particularly when mapping CVEs to threats and vulnerabilities) and a \textit{top\_p} value of 0.9. Furthermore, semantic guardrails and the system context are specified within the prompts to ensure alignment with the CPS architecture, maintain contextual relevance, and minimise speculative content. Specifically, a one-shot example of the expected JSON output format is included in the prompts to enforce schema compliance and guide the LLM towards structured, machine-readable representations of threats and attack paths. Enforcing compliance with an expected JSON schema can be achieved with some LLMs using features such as "structured outputs"~\cite{OpenAIstructured}.

The prompt specifies the persona of a senior cybersecurity expert, in contrast to the senior solution architect persona used during the prior phase of architectural reconstruction. This change reflects the shift in analytical focus from structural composition to adversarial reasoning. Adopting the perspective of a cybersecurity expert guides the model to concentrate on identifying plausible threat actors, enumerating attack surfaces, and formulating structured attack paths consistent with the CPS context. This targeted persona alignment increases the likelihood that the generated threat model and attack tree encapsulate realistic adversarial behaviours while remaining consistent with the previously derived architectural schema.

For the threat model generation, the prompt additionally includes an instruction to produce an "\textit{arch\_suggestions}" field. This field lists  missing or ambiguous architectural information -- such as authentication flows, protocol specifications, safety‑system integration details, or network segmentation data -- that would improve the precision of subsequent threat modelling. This mechanism allows the practitioner to iteratively refine and augment the reconstructed architecture from the previous phase, thereby establishing a feedback loop that enhances both architectural completeness and modelling accuracy.

\begin{figure}[ht]
\centering
\begin{minipage}{0.9\textwidth}
\begin{alltt}\footnotesize
You are a senior cyber security expert with over 20 years of experience in cyber-physical systems (CPS) risk and \\threat modelling, including deep expertise in STRIDE-LM and safety/security co-analysis. You have applied STRIDE-LM \\extensively in ICS, SCADA, and related CPS domains. Your task is to think deeply to thoroughly analyse the provided \\system architectural diagram (e.g., Data Flow Diagram) along with any accompanying documentation to produce a \\comprehensive list of specific threat scenarios relevant to the \\application.

System context: Municipal heating system

Controlled sampling configuration:
- temperature = 0.1
- top_p = 0.9

IMPORTANT - Follow these strictly enforced semantic guardrails:
1. If the diagram includes an "Attacker" entity—whether internal, external, explicit, or implicit—treat it as the origin \\for possible attack paths and enumerate realistic threats accordingly.
2. For each STRIDE-LM category, identify 3 to 4 credible threat scenarios if applicable. Each scenario must describe a \\concrete, context-specific attack, avoiding generic descriptions.
3. Focus your analysis on cyber-physical systems. Address system-level impacts such as disruption of physical processes, \\loss of control, cascading failures, or safety hazards rather than purely IT-centric threats.
4. Consider multiple potential attacker objectives (e.g., power disruption, asset damage, persistent foothold in isolated \\OT environments, bypassing safety controls).
5. Leverage and extract from the accompanying documentation to reflect the assets, vulnerabilities (both CVE-linked and \\non-CVE-linked), hazards, and objectives in each scenario.
6. Identify and list only CVEs that are visible in the accompanying documentation. For each CVE, provide the CVE \\identifier, the affected product, and a brief description. Indicate if the CVE has been observed in known attack campaigns \\(e.g., BlackEnergy, FrostyGoop), with references.
7. Apply FMECA-style reasoning where applicable to identify failure modes, their effects, and potential cascading \\consequences.
8. Format your response strictly as JSON with these top-level keys:
   - `"threat\_model"`: an array of threat scenario objects.
   - `"arch\_suggestions"`: a list of missing architectural information (e.g., authentication flows, protocol details, \\safety system integration, segmentation) needed for more precise modelling.
9. Each threat scenario object must contain the following keys:
   - `"Threat Type"`, based on STRIDE-LM categories (Spoofing, Tampering, Repudiation, Information Disclosure, Denial of \\Service, Elevation of Privilege, Lateral Movement).
   - `"Scenario"`: a detailed narration integrating information about assets, vulnerabilities (including CVE and non-CVE), \\hazards, and attacker objectives. Include references to any CVEs mentioned, and highlight if they were employed in known \\attack campaigns.
   - `"Potential Impact"`
10. Do NOT include general security recommendations or any commentary.
11. Provide no text outside the JSON structure.
\end{alltt}
\end{minipage}
\caption{STRIDE-LM threat model generation prompt with semantic guardrails and controlled sampling.}
\label{fig:LLM-Prompt-ThreatModel}
\end{figure}

\begin{figure}[ht]
\centering
\begin{minipage}{0.9\textwidth}
\begin{alltt}\footnotesize
You are a senior cyber security expert with over 20 years of experience in cyber-physical system (CPS) threat \\management and incident response. Your task is to think deeply to thoroughly analyse the threat model and create an \\attack tree structure in JSON format.

System context: Municipal heating system

Controlled sampling configuration:
- temperature = 0.1
- top_p = 0.9

IMPORTANT - Follow these strictly enforced semantic guardrails:
1. The one and only root node represents the attack goal, which is the disruption or stoppage of cyber-physical system \\operations, taking into account the specific context of the system being analysed.
2. Each node in the tree should represent an Asset, Vulnerability, Hazard, or Goal.
3. The tree should include all relevant attack paths and sub-paths based on the threat model.
4. Analyse if assets, hazards, or vulnerabilities may be linked to assets, hazards, or vulnerabilities in separate \\attack paths, and if so, represent these relationships appropriately in the tree structure.
5. Each node label must begin with a prefix indicating its type:
- `[A##]` for Asset nodes
- `[V##]` for Vulnerability nodes
- `[H##]` for Hazard nodes
- `[G##]` for Goal node(s)
6. Maintain parent-child relationships strictly according to the rules as follows:
- Asset nodes may have children that are Vulnerabilities, Hazards, or other Assets.
- Goal node may have children that are Asset, Vulnerability or Hazard nodes.
- Vulnerability nodes may have children that are Vulnerabilities or Assets, but never Hazards.
- Hazard nodes may have children that are Hazards or Assets, but never Vulnerabilities.
7. The one and only attacker node is at the bottom of the tree structure, connected to all the attack paths leading to the \\attack goal.
- The attacker node should be labelled with the prefix `[U01] Attacker`.
- This attacker node must have children links (edges) to all leaf nodes (the last nodes) in every attack path in the tree.
- This represents the attacker as the origin of all end-stage threats in the attack tree.
8. Use simple IDs (e.g., root, vul1, haz1, asset1).
9. Make labels clear, descriptive, and correctly prefixed.
10. Ensure the JSON is properly formatted. The JSON structure should follow this format:
\{
    "nodes": [
        \{
            "id": "root",
            "label": "Compromise Application",
            "children": [
                \{
                    "id": "auth",
                    "label": "Gain Unauthorised Access",
                    "children": [
                        \{
                            "id": "auth1",
                            "label": "Exploit OAuth2 Vulnerabilities",
                            "children": [
                                \{"id": "attacker", "label": "[U01] Attacker"\}
                            ]
                        \}
                    ]
                \}
            ]
        \}
    ]
\}
\end{alltt}
\end{minipage}
\caption{Attack tree generation prompt with semantic guardrails and controlled sampling.}
\label{fig:LLM-Prompt-AttackTree}
\end{figure}

\subsubsection{Phase 3: Probabilistic Analysis}

The final phase of the workflow constructs a BN model using the architectural reconstruction and security modelling outputs generated in prior phases. This model is then subjected to probabilistic analysis to quantify the probabilities of exposure and severe impact, and the resulting composite risk and availability scores. Following the approach established in our earlier work
~\cite{Huang2026}, probabilistic dependencies among architectural elements, vulnerabilities, and attacker behaviours are modelled within the BN to capture causal relationships and inherent uncertainties. The model is encoded in AutomationML~\cite{AutomationML2025} to provide semantic interoperability and facilitate integration within CPS environments.

Within the BN, nodes represent CPS assets, vulnerabilities, and hazards, while directed edges represent the causal and conditional dependencies between these nodes. This structured representation supports dynamic updates of model attributes, integrating both observed system data and hypothesised attack scenarios. Consequently, it enables the computation of posterior probabilities of attack success, referred to as exposure, and of severe impact to the CPS. The parameterisation of the BN defines the computational logic governing exposure and impact probabilities across vulnerability, asset, and hazard nodes. Vulnerability exposure probabilities are derived from confidence-calibrated, complementary metrics, such as the Common Vulnerability Scoring System (CVSS) \cite{CVSS31} and the Exploit Prediction Scoring System (EPSS) \cite{EPSS}. These are scaled using an \textit{attack feasibility modifier} that accounts for attacker capability and the overall system cybersecurity posture. Asset exposure is treated as a time-dependent failure process, while hazard exposure propagates along cyber-physical dependency chains where upstream disruptions cascade downstream through interconnected control and operational pathways. The computation employs the variable elimination (VE) method \cite{Zhang1996}, enabling posterior risk estimates even for complex, multi-layered architectures. For completeness, full probability formulations are provided in the supplementary material.

\emph{\textbf{Running Example.}} A four-prompt chaining sequence automates the construction and parameterisation of an IEC 62714 (AutomationML)-compliant BN model. This approach leverages prompt chaining to build the model from prior outputs in incremental steps without overwhelming the LLM's context window, by decomposing the overall task into smaller, schema-constrained transformations that can be reliably composed. These steps are:

\begin{itemize}
    \item Constructing XML block structure for asset, vulnerability, and hazard elements.
    \item Extracting directed edges from attack paths from connected element pairs.
    \item Modifying and assigning XML block interfaces to connected element pairs.
    \item Finalising and validating model construction with updated attribute data.
\end{itemize}

These prompts progressively map system context, architectural and threat information, and attribute data into a coherent BN representation. Each prompt operates under a controlled sampling configuration -- with a \textit{temperature} of 0.1 and a \textit{top\_p} value of 0.9 -- alongside semantic guardrails, system context, and one-shot or few-shot examples of the expected schema, thereby ensuring syntactic and semantic compliance of generated BN models suitable for downstream probabilistic analysis. Full prompts are accessible via our public GitHub repository \cite{ASTRAL2026}.

The resulting BN captures key CPS assets, vulnerabilities on those assets, and hazards representing cyber-physical consequences such as full system compromise of the municipal heating system. Directed edges denote dependencies along reconstructed attack paths. Vulnerability nodes are parameterised using the derived \textit{exposure} and \textit{impact} probabilities, while asset nodes incorporate failure‑based exposure models. The attack-goal node aggregates these dependencies to yield posterior probabilities of successful attack and severe impact, corresponding to the two‑day heating outage that affected over 600 apartment buildings in Ukraine during extreme cold. A sample of the probabilistic analysis results, including the resulting composite risk and system availability scores, are presented in Table~\ref{tab:frosty_goop_results}.

The posterior probability of successful attack ($0.3404$) and severe impact ($0.6204$) illustrate how vulnerabilities and hazards within the CPS can propagate across reconstructed cyber-physical dependencies, leading to system-wide disruption. The composite risk score of 21.12\% and system availability of 25.04\% provide a quantitative characterisation of the system’s operational fragility under the modelled attack. Crucially, these estimates are only attainable because the architectural reconstruction phase established the trust boundaries, asset relationships, and attack paths required to parameterise the BN. Without this reconstruction step, such quantitative risk reasoning would not have been feasible in the presence of incomplete documentation.

\begin{table}[ht]
    \centering
    \footnotesize
    \caption{Sample Probabilistic Analysis Results: FrostyGoop (21.12\% Risk Score, 25.04\% System Availability)}
    \begin{tabular}{ll}
    \toprule
    \textbf{Metric} & \textbf{Value} \\
    \midrule
    Sample Element ID & V05 (Vulnerability Node) \\
    Description & Exploit Weak HMI Authentication or Session Hijacking \\
    CVE ID & No CVE Identifier \\
    CVSS Vector & AV:N/AC:L/PR:N/UI:N/S:U/C:H/I:H/A:N \\
    Base $P(\text{Exposure})$ & 0.47 \\
    Calibrated $P(\text{Exposure})$ & Mean=0.4982, 95\% CI=(0.4032, 0.5933) \\
    Updated $P(\text{Exposure})$ & 0.4982 \\
    $P(\text{Severe Impact})$ & 0.45 \\
    \midrule
    Sample Element ID & A02 (Asset Node) \\
    Description & Wireless Field Devices (Sensors/Actuators) \\
    $P(\text{Exposure})$ & 0.4646 \\
    $P(\text{Severe Impact})$ & 0.0159 \\
    \midrule
    Sample Element ID & H20 (Hazard Node) \\
    Description & Full Compromise of Heating Control System \\
    $P(\text{Exposure})$ & 1.00 \\
    $P(\text{Severe Impact})$ & 0.0159 \\
    \midrule
    \textbf{$P(\text{Successful Attack})$} & \textbf{0.3404} \\
    \textbf{$P(\text{Severe Impact})$} & \textbf{0.6204} \\
    \textbf{Risk Score} & \textbf{21.12\%} \\
    \textbf{System Availability} & \textbf{25.04\%} \\
    \bottomrule
    \end{tabular}
    \label{tab:frosty_goop_results}
\end{table}

\subsection{Continuous Architectural Refinement}

The ASTRAL framework enables the dynamic computation of contextually relevant security risk and system availability metrics, thereby supporting a continuous and adaptive CPS security assessment process. By updating the BN model through configurable \textit{countermeasure portfolios}, i.e., combinations of vulnerability mitigation attributes, and real-time threat and system data, the system supports real-time simulation of alternative countermeasure configurations and adaptation to evolving incidents. This capability assists practitioners in prioritising remediation actions in line with evolving threat conditions and specific operational objectives, such as ensuring that the CPS remains available to deliver essential services even when under attack. For instance, a practitioner can use ASTRAL to simulate and select an appropriate countermeasure portfolio that reduces overall risk whilst ensuring the system remains available, by adjusting mitigation attribute values for selected vulnerability nodes within the model. 

Furthermore, the integration of the LLM with security assessment facilitates continuous architectural refinement through iterative prompting and real-time data updates. Practitioners can incorporate updated metadata as it becomes available, thereby systematically reducing architectural incompleteness over time. As newly discovered threat, attack, and system details are fed back into the LLM, the architectural narration is refined, which in turn triggers an updated parameterisation of the BN. This closed-loop process helps to ensure that the security assessment remains accurate even as the threat landscape evolves or more granular documentation is uncovered.

\section{CPS Case Studies}
\label{sec:case_studies}

Beyond the FrostyGoop case study used to explain our approach, we further evaluate the applicability of ASTRAL across other CPSs that feature contrasting architectural characteristics and documentation gaps.

\paragraph{Medical Cyber-Physical Systems (MCPSs)} In the context of MCPSs~\cite{Zhang2015, Chen2021}, ASTRAL can ingest artefacts such as high‑level schematic diagrams that describe body‑area sensors, software‑defined networking (SDN) controllers and gateways, cloud‑based medical data management platforms, and Electronic Health Record (EHR) systems. The LLM reconstructs trust boundaries between architectural elements and further identifies critical interfaces such as SDN routers, wireless links, and API endpoints, thereby supporting threat and attack modelling scenarios. This includes scenarios in which compromised wearables or gateways may be exploited to manipulate patient data, disrupt clinical workflows, or exfiltrate sensitive medical records.

\begin{table*}[ht]
\centering
\footnotesize
\caption{Sample Probabilistic Analysis Results (MCPS: 39.15\% Risk Score, 14.01\% System Availability | Solar PV: 4.95\% Risk Score, 38.65\% System Availability)}
\label{tab:probabilistic_results}
\begin{tabular}{ll|ll}
\toprule
\multicolumn{2}{c|}{\textbf{Medical CPS (MCPS)}} & \multicolumn{2}{c}{\textbf{Solar PV Inverter}} \\
\midrule
\textbf{Metric} & \textbf{Value} & \textbf{Metric} & \textbf{Value} \\
\midrule
Sample Element ID & V01 (Vulnerability Node) & Sample Element ID & V05 (Vulnerability Node) \\
Description & Wind River VxWorks Buffer Overflow & Description & Physical Access to Inverter Configuration Ports \\
CVE ID & CVE-2019-12255 & CVE ID & No CVE Identifier \\
CVSS Vector & AV:N/AC:L/PR:N/UI:N/S:U/C:H/I:H & Proxy-CVSS Vector & AV:P/AC:L/PR:N/UI:N/S:U/C:H/I:H \\
$P(\text{Exposure})$ & 0.9 & Base $P(\text{Exposure})$ & 0.11 \\
 & & Calibrated $P(\text{Exposure})$ & Mean=0.4771, 95\% CI=(0.3820, 0.5721) \\
 & & Updated $P(\text{Exposure})$ & 0.4771 \\
$P(\text{Severe Impact})$ & 0.85 & $P(\text{Severe Impact})$ & 1.0 \\
\midrule
Sample Element ID & A03 (Asset Node) & Sample Element ID & A29 (Asset Node) \\
Description & EHR System & Description & Inverter Safety Settings \\
$P(\text{Exposure})$ & 0.2683 & $P(\text{Exposure})$ & 0.0308 \\
$P(\text{Severe Impact})$ & 0.003 & $P(\text{Severe Impact})$ & 0.0002 \\
\midrule
Sample Element ID & H06 (Hazard Node) & Sample Element ID & H06 (Hazard Node) \\
Description & Exfiltration of medical imaging data & Description & Physical Threats to Inverter or Field Devices \\
$P(\text{Exposure})$ & 1.00 & $P(\text{Exposure})$ & 1.00 \\
$P(\text{Severe Impact})$ & 0.003 & $P(\text{Severe Impact})$ & 0.0002 \\
\midrule
\textbf{$P(\text{Successful Attack})$} & \textbf{0.6214} & \textbf{$P(\text{Successful Attack})$} & \textbf{0.0858} \\
\textbf{$P(\text{Severe Impact})$} & \textbf{0.6300} & \textbf{$P(\text{Severe Impact})$} & \textbf{0.5772} \\
\textbf{Risk Score} & \textbf{39.15\%} & \textbf{Risk Score} & \textbf{4.95\%} \\
\textbf{System Availability} & \textbf{14.01\%} & \textbf{System Availability} & \textbf{38.65\%} \\
\bottomrule
\end{tabular}
\end{table*}

Sample probabilistic analysis results for a representative MCPS~(Table~\ref{tab:probabilistic_results}) indicate a moderate risk score of 39.15\% and a low system availability of 14.01\%. The risk score reflects the overall likelihood that exploitable vulnerabilities, such as gateway buffer overflows or insecure wireless channels, could lead to severe consequences affecting patient safety or data integrity across interconnected system layers. In contrast, the low availability value quantifies the expected operational degradation once those attack pathways materialise, indicating that a successful compromise would quickly disable or disrupt a substantial portion of clinical functions. These figures underscore the inherently high‑risk nature of medical CPS environments, where interconnected wearables, edge gateways, and cloud services form dense dependency chains capable of propagating failures. Through ASTRAL’s reconstructed architectural model -- clarifying hidden trust boundaries and unobserved API dependencies --practitioners gain a precise understanding of both the probability of compromise (captured by the risk score) and the extent of impact on service continuity (captured by availability), enabling prioritised mitigation of vulnerabilities within MCPSs.

\paragraph{Distributed Solar Power Installations} In the energy domain, ASTRAL can also be applied to distributed solar power installations using a BN graph derived from partial site diagrams, inverter‑network schematics, and vendor advisories, such as those reported in the SUN:DOWN study \cite{Forescout2025} on solar photovoltaic (PV) inverter vulnerabilities. From these heterogeneous artefacts, ASTRAL produces an architectural narration that captures the relationships among field‑level inverters, local controllers, aggregation gateways, and grid‑facing interfaces, including remote‑management channels. This representation facilitates LLM‑driven identification of coordinated attack paths that may target multiple inverters simultaneously and supports Bayesian risk analysis quantifying hazards such as localised grid instability or loss of generation capacity under orchestrated cyber‑physical attacks~\cite{Dabrowski2017}.

Sample probabilistic analysis results for a representative solar PV inverter system (Table~\ref{tab:probabilistic_results}) indicate a very low risk score of 4.95\% and a comparatively higher system availability of 38.65\%. The low risk score indicates that the probability of severe cyber‑physical compromise across the entire inverter network remains limited, reflecting the relatively contained exposure surface and predominantly physical‑domain vulnerabilities. In contrast, the higher availability value denotes that, even under plausible attack conditions, core energy generation and control functions would remain largely operational, with only partial or localised disruptions. These outcomes align with the loosely coupled architecture of distributed solar installations, where individual inverter faults seldom cascade to the grid level. By inferring missing architectural details -- such as remote‑management pathways and field‑to‑SCADA interfaces -- ASTRAL enables practitioners to visualise credible multi‑point attack chains and quantitatively estimate their local versus systemic impact. The resulting insight supports targeted mitigation planning around inverter configuration access controls and field‑device hardening, ensuring that mitigation measures are proportional to measured exposure and availability risks.

Taken together with the FrostyGoop running example, these case studies (accessible via our public GitHub repository \cite{ASTRAL2026}) demonstrate that ASTRAL’s architecture‑centric workflow generalises effectively across diverse CPS domains, each exhibiting different patterns of architectural incompleteness and operational constraint. To highlight the specific nature of architectural incompleteness encountered across domains, and the extent to which ASTRAL’s LLM‑driven reconstruction improved analytical fidelity, Table~\ref{tab:astral_reconstruction_summary} compares the pre‑ and post‑analysis states for the MCPS and solar PV inverter case studies. The table summarises initially missing architectural elements, how the LLM completed or inferred them from available artefacts, and the tangible analytical gains practitioners can achieve.

\begin{table}[ht]
\centering
\footnotesize
\caption{Summary of Incomplete Artefacts and ASTRAL Reconstruction Effects across CPS Case Studies}
\begin{tabular}{p{2.2cm}p{4.8cm}p{6cm}}
\toprule
\textbf{Case} & \textbf{Missing Artefact Elements} & \textbf{ASTRAL Reconstruction \& Practitioner Gain} \\
\midrule
\textbf{FrostyGoop (Incomplete DFD)} & 
• Missing HMI/PLC authentication mechanisms &
• Reconstructed HMI/PLC authentication flows across 5 trust boundaries (client$\to$API$\to$DB$\to$processing$\to$export) \\
& 
• No specific communication protocols or network segmentation strategies &
• Added protocol details and network segmentation for 7 new attack paths (API key leakage, privilege chaining) \\
& 
• Absent safety system integration and monitoring/logging mechanisms &
• \textbf{Practitioner Gain}: Complete threat model with safety integration and monitoring from partial DFD \\

\midrule
\textbf{MCPS (High-level Schematic)} & 
• Missing detailed authentication/authorisation flows &
• Reconstructed authentication flows across 4 trust boundaries (wearable$\to$gateway$\to$cloud$\to$EHR) \\
& 
• No specific communication protocols or network segmentation details &
• Added protocol mappings and segmentation for 9 attack paths (API injection, cross-domain abuse) \\
& 
• Absent safety system integration and physical-cybersecurity controls &
• \textbf{Practitioner Gain}: Exposed safety dependencies and physical-cyber controls \\

\midrule
\textbf{Solar PV Inverter (BN Graph)} & 
• Missing HMI/PLC/RTU authentication flows &
• Reconstructed HMI/PLC/RTU authentication across 3 trust boundaries (field$\to$aggregator$\to$cloud$\to$SCADA) \\
& 
• No protocol details or SCADA network segmentation &
• Added protocol details and SCADA segmentation for 5 attack paths (remote auth bypass) \\
& 
• Absent safety system integration and firmware update mechanisms &
• \textbf{Practitioner Gain}: Physical-cyber coupling with firmware security controls \\
\bottomrule
\end{tabular}
\label{tab:astral_reconstruction_summary}
\end{table}

\section{Ablation Study}
\label{sec:Ablation_Study}

To address the research question RQ1: \textit{"To what extent do ASTRAL's key design elements contribute to architectural reconstruction quality and structural validity?"}, we conducted ablation experiments using \emph{FrostyGoop} (our running example from Section~\ref{sec:Methodology} with incomplete DFD), \emph{Medical CPS (MCPS)} (high-level schematic), and \emph{Solar PV Inverter} (BN graph) from Section~\ref{sec:case_studies}.  Four variants systematically disable or modify key design elements relative to a \emph{Full ASTRAL} baseline (multimodal input with guardrails enabled, temperature~=~0.1, top-$p$~=~0.9): (1) \emph{Text-Only} (no diagrams, guardrails ON), (2) \emph{No-Guardrails} (full multimodal, guardrails OFF), (3) \emph{High Temperature} (multimodal+guardrails, temperature~=~0.9), and (4) \emph{Low top-$p$} (multimodal+guardrails, top-$p$~=~0.1). Each variant was run ten times per case study using identical prompts and the same LLM (Mistral 3.1 Medium). We evaluate reconstruction quality across four objective, no-ground-truth metrics capturing architectural reconstruction accuracy.

\begin{itemize}
\item \emph{Valid Trust Boundaries}: The count of identified security perimeters that respect documented communication protocols and logical isolation (e.g., wireless sensor to gateway transitions, rather than direct EHR to wearable interconnections).
\item \emph{Graph Connectivity}: The ratio of network nodes successfully reachable from documented entry points, representing the structural integrity of the reconstructed CPS topology.
\item \emph{Impossible Attack Paths}: The count of generated attack paths that violate physical or logical network segmentation (e.g., a direct connection from a field-level inverter to a SCADA system without passing through an intermediary aggregator or gateway).
\item \emph{Semantic Coherence}: An \textit{LLM-as-a-judge}~\cite{Zheng2023judging} metric quantifying the logical consistency and technical clarity of the reconstructed architecture. The scoring prompt directs an independent LLM to act as an impartial judge: ``Please act as an impartial judge and evaluate how semantically coherent is this architecture description for CPS security analysis? Be as objective as possible. Please rate on a Likert scale of 1-5.''
\end{itemize}

\begin{table*}[ht]
\centering
\caption{Summary: Ablation Study Results Across Three Case Studies (Mean $\pm$ SD, $n=10$)}
\label{tab:vertical_ablation_summary}
\footnotesize
\newcolumntype{Y}{>{\centering\arraybackslash}X}
\begin{tabularx}{\textwidth}{l l YYYY}
\toprule
\textbf{Case Study} & \textbf{Variant} & \textbf{Trust BD} & \textbf{Conn.} & \textbf{Impos.} & \textbf{Coherence} \\
\midrule
\multirow{5}{*}{\textbf{FrostyGoop}} 
 & Full ASTRAL   & 5.10 $\pm$ 1.29 & 1.00 $\pm$ 0.00 & 0.00 $\pm$ 0.00 & 4.85 $\pm$ 0.05 \\
 & Text-Only     & 3.90$^\ast$ $\pm$ 0.88 & 1.00 $\pm$ 0.00 & 0.00 $\pm$ 0.00 & 4.58$^\ast$ $\pm$ 0.14 \\
 & No-Guardrails & 4.00$^\ast$ $\pm$ 1.05 & -- & -- & 4.89$^\ast$ $\pm$ 0.03 \\
 & High Temp     & 7.50$^\ast$ $\pm$ 1.08 & 1.00 $\pm$ 0.00 & 0.00 $\pm$ 0.00 & 4.64$^\ast$ $\pm$ 0.18 \\
 & Low top-$p$   & 6.70$^\ast$ $\pm$ 1.42 & 1.00 $\pm$ 0.00 & 0.00 $\pm$ 0.00 & 4.75$^\ast$ $\pm$ 0.16 \\
\midrule
\multirow{5}{*}{\textbf{Medical CPS}} 
 & Full ASTRAL   & 4.80 $\pm$ 1.32 & 1.00 $\pm$ 0.00 & 0.00 $\pm$ 0.00 & 4.84 $\pm$ 0.07 \\
 & Text-Only     & 4.10 $\pm$ 1.29 & 1.00 $\pm$ 0.00 & 0.00 $\pm$ 0.00 & 4.69$^\ast$ $\pm$ 0.14 \\
 & No-Guardrails & 4.30 $\pm$ 1.42 & -- & -- & 4.89 $\pm$ 0.10 \\
 & High Temp     & 7.90$^\ast$ $\pm$ 1.60 & 1.00 $\pm$ 0.00 & 0.00 $\pm$ 0.00 & 4.82 $\pm$ 0.04 \\
 & Low top-$p$   & 8.60$^\ast$ $\pm$ 1.65 & 1.00 $\pm$ 0.00 & 0.00 $\pm$ 0.00 & 4.83 $\pm$ 0.08 \\
\midrule
\multirow{5}{*}{\textbf{Solar PV Inverter}} 
 & Full ASTRAL   & 6.50 $\pm$ 1.65 & 1.00 $\pm$ 0.00 & 0.00 $\pm$ 0.00 & 4.73 $\pm$ 0.13 \\
 & Text-Only     & 3.70$^\ast$ $\pm$ 0.95 & 1.00 $\pm$ 0.00 & 0.00 $\pm$ 0.00 & 4.58$^\ast$ $\pm$ 0.18 \\
 & No-Guardrails & 4.70$^\ast$ $\pm$ 1.49 & -- & -- & 4.81$^\ast$ $\pm$ 0.07 \\
 & High Temp     & 6.70 $\pm$ 1.77 & 1.00 $\pm$ 0.00 & 0.00 $\pm$ 0.00 & 4.74 $\pm$ 0.13 \\
 & Low top-$p$   & 5.40 $\pm$ 1.43 & 1.00 $\pm$ 0.00 & 0.00 $\pm$ 0.00 & 4.00$^\ast$ $\pm$ 0.97 \\
\bottomrule
\addlinespace[0.2cm]
\multicolumn{6}{p{0.95\textwidth}}{\scriptsize \emph{Legend:} \textbf{Trust BD} = valid trust boundaries; \textbf{Conn.} = graph connectivity; \textbf{Impos.} = impossible attack paths; \textbf{Coherence} = semantic coherence (scored 1--5). Results are reported as Mean $\pm$ SD across $n=10$ experimental runs. The $^\ast$ denotes statistical significance ($p \le 0.10$) using the Wilcoxon signed-rank test against the Full ASTRAL baseline (multimodal + guardrails + low temp + high top-$p$).} \\
\end{tabularx}
\end{table*}

\subsection{Key Findings and Discussion}

The results of our ablation study, detailed in Table~\ref{tab:vertical_ablation_summary}, demonstrate several findings regarding CPS architectural reconstruction using ASTRAL.

\textbf{Text-Only Ablation.} Text-only inputs consistently degrade performance across all case studies. Valid Trust Boundaries (Trust BD) were significantly reduced in FrostyGoop ($-23.5\%$), MCPS ($-14.6\%$), and most notably in the Solar PV Inverter ($-43.1\%$). These results confirm the necessity of architecture diagrams for CPS topology reconstruction.

\textbf{No-Guardrails Ablation.} The removal of guardrails led to a decrease in Trust BD for Solar PV Inverters ($-27.7\%$). While the quantitative drop in FrostyGoop ($-11.6\%$) and MCPS ($-10.4\%$) were more moderate, variants without guardrails qualitatively fail to generate usable attack models across all artefacts, thereby preventing downstream threat and attack modelling.

\textbf{Sampling Parameter Ablations.} High Temperature (0.9) and Low top-$p$ (0.1) configurations generally increased the identification of valid Trust Boundaries, suggesting that stochasticity helps the LLM overcome baseline biases. FrostyGoop saw increases up to $+47\%$, while MCPS exhibited the most dramatic gains, increasing Trust BD by up to $+79.2\%$.

\textbf{Artefact Topology and Influence Effects.} The varying sensitivity across case studies reflects the structural clarity and density of the input diagrams:
\begin{itemize}
    \item \emph{Solar PV Inverter (Bayesian Network):} The clear-cut directed edges of the BN graph explicitly define node-to-node influence. This structural clarity likely explains why this case study was the most resistant to sampling-induced fluctuations in Trust BD, as the valid paths were topologically self-evident.
    \item \emph{FrostyGoop (Partial DFD):} As this DFD represented only a localised subset of a larger system, it acted as a constrained topological anchor. The model was restricted by the specific data flows provided, leading to moderate sensitivity.
    \item \emph{Medical CPS (High-Level Schematic):} The abstract nature of this schematic induced the highest sensitivity. The lack of explicit influence paths (compared to the Solar PV BN) allowed the LLM to utilise the increased creative freedom of stochastic sampling to fill in implied architectural gaps, resulting in the largest numerical gains in Trust BD.
\end{itemize}

\textbf{Error Suppression and Robustness.} Across all three case studies, the pipeline demonstrated remarkable structural robustness. Connectivity (Conn.) remained at a perfect $1.00$ and Impossible Paths (Impos.) remained at $0.00$ for every variant where guardrails were active. Even under perturbed sampling or abstract inputs, the ASTRAL framework successfully suppressed topologically invalid connections.

\begin{tcolorbox}[width=1.0\textwidth, colback=gray!7!white, colframe=black, boxrule=0.6pt, arc=4pt, leftrule=0pt, rightrule=0pt, top=1pt, bottom=1pt]
\small
\textbf{RQ1 Key Findings}: The ablation study validates ASTRAL’s design. Explicit influence graphs (like the Solar PV BN) provide the strongest grounding, while abstract schematics (MCPS) benefit most from controlled stochasticity to uncover latent boundaries. In all cases, architecture diagrams are essential, and guardrails are the primary mechanism for maintaining structural integrity.
\end{tcolorbox}

\section{Validation by Practitioners}
\label{sec:Validation_From_Practitioners}

In this section, we aim to answer the research questions RQ2, RQ3, and RQ4, by gathering cybersecurity practitioners' perspectives on ASTRAL's usefulness, trustworthiness, and reliability. A mixed-methods research design, incorporating both quantitative and qualitative data from a selected participant group, allowed for comprehensive validation of our approach.

\subsection{Validation Methodology}

Our validation followed a structured methodology, combining guided demonstration sessions with practitioner surveys, consistent with recognised practices for cybersecurity tool evaluation \cite{Wijayarathna2018}. Fourteen participants were selected from the first author's professional network, comprising Chief Information Security Officers (CISOs) and senior cybersecurity practitioners with CPS domain expertise and practical experience conducting architecture-centric security assessments. These roles represent the primary intended users of ASTRAL, making their input essential for validating the tool's real-world applicability. Following established expert evaluation protocols in cybersecurity research mitigated selection bias and ensured the relevance of practitioner feedback \cite{Bernsmed2022}.

The validation process consists of two phases. In the first phase, participants received an individual demonstration of the prototype tool, delivered either in-person or online. During these sessions, participants were guided through the key functionalities, including the sequential threat risk assessment workflow illustrated in Figure~\ref{fig:ASTRAL_Overview}. A recorded version of the demonstration given to each participant is available in our repository~\cite{ASTRAL2026}. In the second phase, participants were given independent access to the tool after the demonstration to evaluate its functionality and assess its practical usefulness and limitations. Participants were then invited to complete a survey and provide their feedback. For consistency, in-person sessions used a standard laptop supplied by the first author. However, all participants ultimately selected the online format and received web-based access links, minimising technical barriers to participation.

\subsection{Survey Instrument Design}

The survey used a mixed-methods approach, combining quantitative Likert-scale ratings (1--5 scale) with qualitative open-ended questions for detailed insights. This design enables both statistical analysis and the collection of qualitative feedback to refine ASTRAL. The survey comprised nine questions, as listed below:

\begin{enumerate}
    \item What is your current role in the field of cybersecurity?
    \item To what extent do you find the LLM-powered threat risk assessment tool useful? (1: Not at all useful – 5: Extremely useful)
    \item To what extent do you consider the output from the LLM-powered threat risk assessment tool to be trustworthy and reliable? (1: Not at all reliable – 5: Extremely reliable)
    \item How would you rate your overall experience in using and navigating the tool? (1: Very difficult – 5: Very easy)
    \item How likely are you to use or recommend this tool in your professional work? (1: Not at all likely – 5: Extremely likely)
    \item What key strengths have you experienced when using the tool?
    \item What limitations or challenges have you encountered while using the tool?
    \item What improvements would you suggest to enhance the tool’s usefulness and reliability?
    \item Is there anything else you would like to share about your experience with the tool?
\end{enumerate}

The administrative configuration of the survey was such that participation was voluntary and anonymous; no login was required to access the tool, and no interaction data was recorded during the sessions.

\subsection{Results}

The survey results presented in Table~\ref{tab:participant_responses} indicate that participants rated ASTRAL highly across key metrics, including usefulness, ease of use, trustworthiness and reliability, and likelihood of future use or recommendation. Among these metrics, usefulness received the highest average rating (4.64), whereas trustworthiness and reliability of the outputs were rated lowest (4.00).

\begin{table}[ht]
\centering
\footnotesize
\caption{Average participant ratings (Likert scale: 1--5)}
\label{tab:participant_responses}
\begin{tabular}{p{6cm} c}
\toprule
\textbf{Evaluation Criterion} & \textbf{Average Rating} \\
\midrule
Usefulness of the tool & 4.64 \\
Ease of use and navigation & 4.29 \\
Likelihood of future use or recommendation & 4.36 \\
Trustworthiness and reliability of outputs & 4.00 \\
\bottomrule
\end{tabular}
\end{table}

Participants offered diverse feedback regarding ASTRAL's strengths, limitations, and avenues for further improvement. We address the research questions RQ2, RQ3, and RQ4 by synthesising this feedback.\\

\textbf{RQ2: What architectural information gaps can ASTRAL fill?}
The feedback highlights ASTRAL’s capability to automatically extract and structure architectural details from diagrams and textual descriptions, efficiently identifying system components, trust boundaries, and potential attack points. One participant noted, \textit{“What stood out most to me was how quickly the tool could generate a structured view of the system just from an architectural diagram. It identified the key components, trust boundaries, and attack entry points quite intuitively -- the kind of groundwork that usually takes time to piece together manually.”}

ASTRAL's interpretation of architecture diagrams with minimal manual input was considered particularly valuable for complex systems with incomplete or unavailable documentation. Features such as statistical risk assessment, attacker path modelling, and attack paths were noted to contribute to comprehensive asset categorisation and help address architectural information gaps. However, the completeness, relevance, and practical value of outputs remain closely dependent on the quality and specificity of user-supplied artefacts. When architectural diagrams or accompanying documents fail to accurately represent the underlying system, the resulting analysis tends to yield generalised conclusions, limiting its usefulness for case-specific modelling and decision-making. Participants therefore emphasised increased flexibility for uploading supplementary artefacts, such as configuration files, inventory lists, and additional system data, to enable more nuanced and context-aware assessments.

Deeper integration with live system inventories and operational databases was identified as essential to reflect real-time asset exposure and tailor risk evaluations. Enhancing interactive prompts to better capture operational context and constraints would further strengthen scenario-specific threat analyses and the practical relevance of architecture-centric security assessments. This would reinforce ASTRAL’s ability to provide contextually precise insights and address gaps in architectural documentation using LLMs.

\begin{center}
\begin{tcolorbox}[width=1.0\textwidth, colback=gray!7!white, colframe=black, boxrule=0.6pt, arc=4pt, leftrule=0pt, rightrule=0pt, top=1pt, bottom=1pt]
\textbf{RQ2 Key Findings}:
\small
ASTRAL effectively mitigates incomplete architectural knowledge through multimodal LLM techniques. Survey participants recognised its usefulness in synthesising architectural insights with minimal manual effort. Expanding support for supplementary artefacts, live operational data, and interactive context capture can further enhance the depth, accuracy, and contextual relevance of its assessments.
\end{tcolorbox}
\end{center}

\textbf{RQ3: How do cybersecurity practitioners perceive the trustworthiness and reliability of the security assessment results generated by ASTRAL?}
Participants generally considered ASTRAL to be trustworthy and reliable, as reflected in an average rating of 4.00 for “Trustworthiness and reliability of outputs”. ASTRAL’s systematic and repeatable methodology, grounded in recognised cybersecurity frameworks, enabled participants to validate assessment findings and substantiate conclusions with confidence. The tool was perceived as effective for preliminary evaluations and for the prompt verification and comparison of other security assessment outcomes, including those from audits and vulnerability scans. As one participant remarked, \textit{“It is a strong ‘first-cut’ tool that helps kick-start threat and risk assessments. Instead of starting from scratch or referencing old templates, this gives a quick and structured baseline to work from, especially useful when assessments are a regular part of the work.”}

Despite the positive feedback, participants identified several avenues to further improve the trustworthiness and reliability of ASTRAL’s outputs. Greater transparency, explainability and traceability were considered essential for professional scrutiny and benchmarking, particularly in relation to STRIDE-LM threat assessments. One participant recommended referencing real-world cases or well-established scenarios to justify each identified threat or risk score, thereby strengthening confidence in the system’s underlying reasoning. Additionally, participants proposed that supplementary guidance and more detailed explanations of assessment outputs would further reinforce trust in the system’s recommendations.

Enhancing interactivity and guided user feedback was widely recognised as a mechanism to improve reliability and practitioner confidence. Although users can already input contextual information, the introduction of additional guidance during this step could make the process more structured. Prompts encouraging users to more clearly specify assumptions and operational constraints would yield outputs that are more relevant and contextually accurate. As one participant commented, \textit{“A guided context input step would be valuable. Prompts that nudge users to describe the system’s purpose, operating assumptions, or known constraints could make the generated threats and risk scores more accurate. It’s often this operational context that determines whether a threat is theoretical or genuinely high-risk.”}

Several participants further recommended “human-in-the-loop” interventions, enabling practitioners to refine model outputs or override underlying assumptions. As one participant noted, \textit{“It might be worth exploring a way for users to fine-tune or override certain assumptions. For example, being able to adjust the perceived impact or exploitability after reviewing the threat assessment results would give practitioners more control while still leveraging the automation.”}

\begin{center}
\begin{tcolorbox}[width=1.0\textwidth, colback=gray!7!white, colframe=black, boxrule=0.6pt, arc=4pt, leftrule=0pt, rightrule=0pt, top=1pt, bottom=1pt]
\small
\textbf{RQ3 Key Findings}:\
ASTRAL is perceived as a trustworthy and reliable platform for CPS security assessment, providing structured, professional outputs aligned with recognised cybersecurity frameworks. Enhancing transparency, explainability, and traceability -- together with guided context capture, human-in-the-loop interactions, and configurable parameters -- will further strengthen trust and improve the reliability of LLM-driven security assessments.
\end{tcolorbox}
\end{center}

\textbf{RQ4: How useful is our proposed approach to CPS practitioners?}
Participant feedback strongly supports the practical usefulness of ASTRAL, as reflected by high average ratings for “Usefulness of the tool” (4.64) and “Likelihood of future use or recommendation” (4.36). Participants emphasised the efficiency gains enabled by ASTRAL’s intuitive and structured sequential workflow. Its integrated design was regarded as instrumental in streamlining security assessments and promoting collaboration between cybersecurity and IT teams. The capability to conduct comprehensive security assessments without the need to alternate between multiple tools or frameworks was consistently highlighted as a significant advantage for professional adoption. As one participant remarked, \textit{“I liked how the workflow links each step together. It makes the process feel coherent and guided, rather than having to switch across different tools or frameworks.”}

Participants particularly valued ASTRAL’s ability to generate architecture-driven insights and structured outputs, which facilitate the prioritisation of defensive actions. Statistical risk quantification and contextual mapping of attack paths were recognised for their operational relevance, supporting expedited triage and decision-making. Participants further noted ASTRAL’s adaptability to the requirements of diverse cybersecurity roles and multidisciplinary teams. Its capacity for contextual adaptation, such as permitting the incorporation of organisation-specific assumptions and operational constraints, was seen to enhance the quality and relevance of its security assessments.

The flexibility of ASTRAL was viewed positively, particularly its support for different LLM providers and models, its quantitative approach to risk scoring, and its guidance towards actionable security recommendations. The tool’s alignment with secure-by-design and risk-based assessment principles was regarded as especially valuable for organisations managing complex or evolving systems. As one participant observed: \textit{“It provided very useful information to understand the components in a typical OT architecture diagram, including threats, risks, and attack paths, which saves considerable time otherwise spent collecting information from multiple sources.”}

Several participants proposed improvements to further increase ASTRAL’s practical usefulness. A key suggestion was to enable users to specify known or anticipated threat actors relevant to their environment, thereby allowing the system to tailor its threat modelling and risk evaluations to reflect actual adversaries and tactics in their operational contexts. As one participant explained: \textit{“By allowing users to specify known or anticipated threat actors relevant to their environment, the tool could tailor its threat modelling and risk evaluation to reflect real-world adversaries and tactics. This targeted approach would enhance the accuracy of risk assessments and ensure that mitigation strategies are aligned with the most pertinent threat scenarios facing the organisation.”}

Participants further recommended the incorporation of mechanisms for dynamically capturing mitigation controls and updating risk profiles as system configurations or the threat landscape evolve. Automating attack feasibility assessments through integration with operational data and threat intelligence frameworks was identified as another valuable enhancement. One participant suggested: \textit{“It would be useful to take into account actual risk and vulnerability data from scanning results and asset databases, to reflect the real-life applicability of attack paths more accurately - such as whether specific assets and versions are susceptible to given attack scenarios. This would help users prioritise remediation efforts more effectively.”} Other suggested improvement areas include: deeper integration with established cybersecurity frameworks, advanced data visualisation capabilities (such as spider charts or overlays connecting threats to architecture diagrams), support for additional commercial LLM providers, and expanded export and output customisation options. These improvement areas were regarded as essential for maximising ASTRAL’s analytical robustness, transparency and long-term relevance in professional settings.

\begin{center}
\begin{tcolorbox}[width=1.0\textwidth, colback=gray!7!white, colframe=black, boxrule=0.6pt, arc=4pt, leftrule=0pt, rightrule=0pt, top=1pt, bottom=1pt]
\small
\textbf{RQ4 Key Findings}:\
ASTRAL is rated highly for usefulness, ease of use, and anticipated adoption, reinforcing its professional utility. Survey participants identified opportunities to enhance its practical value through dynamic feedback capture, richer data integration, and improved data visualisation. Further enhancements, including threat actor modelling, automated attack feasibility assessment, and expanded reporting options, are expected to strengthen its relevance and practitioner acceptance.
\end{tcolorbox}
\end{center}

\subsection{Overall Assessment}

The survey results indicate that ASTRAL provides substantial benefits for architecture-centric security assessments in CPS environments. Participants reported high levels of satisfaction with its ease of use, structured analytical workflow, and its ability to streamline resource-intensive security assessment processes.

The incorporation of advanced capabilities, such as automated extraction of architectural information, multimodal LLM reasoning, Bayesian model analysis, probabilistic risk scoring, and automated derivation of attack paths, positions ASTRAL as an effective resource for both preliminary security assessment and cross-validation of threat and risk analyses. Its adaptability across diverse practitioner roles and operational contexts also facilitates collaborative assessments among stakeholders and teams, thereby promoting best practice in cybersecurity management. In addition, ASTRAL’s potential to generate synthetic CPS data where real data is unavailable, inaccurate, or inadequate enables more robust and realistic security assessments under constrained conditions. As one participant remarked: \textit{“Overall, I think it is an excellent idea and effort. Truth be told, while I know this is theoretically possible, seeing the tool actually in action with reasonably good output has helped trigger further thoughts on how the tool can be used in different circumstances. I would be keen to explore these further.”}

Although participants acknowledged ASTRAL’s strengths, a number of areas for further enhancement were identified to address its limitations and bolster practical relevance. Some participants observed inconsistencies when deploying ASTRAL across diverse datasets and assessment scenarios, which affected perceptions of its trustworthiness and reliability. As non-deterministic outcomes are inherent to LLM-based approaches, it is expected that different outputs may be generated with each execution. Furthermore, the risk of hallucinations can lead to spurious or unverifiable outputs. Future improvements should therefore focus on advancing LLM parameter tuning and providing users with clearer explanations regarding the underlying causes of output variability.

Performance concerns were also highlighted, such as delays encountered during attack path generation, especially in complex architectural situations. Adoption of commercial LLM services with higher-performance models may help to mitigate these latency issues. Moreover, while selection bias persists as a potential limitation, the exploratory scope of this evaluation emphasises the importance of situating these findings within context. Additional limitations highlighted by the participants include: limited control over the initial configuration of security assessments, information overload, and suboptimal data visualisation in the tool. The absence of human-in-the-loop parameter tuning and the inability to map threats directly onto architectural diagrams were also noted as constraints impacting the utility of assessments. In response, participants recommended measures to improve transparency and interpretability of outputs, facilitate continuous feedback and dynamic reassessment, and enable deeper integration with live asset inventories and established industry frameworks. Enhancing data visualisation and export capabilities was also suggested to advance analytical clarity and reporting flexibility. Collectively, implementing these improvements would enhance ASTRAL's practical usefulness, trustworthiness, and reliability.

\subsection{Threats to Validity and Future Work}

While practitioner feedback was positive, several threats to validity~\cite{Cruzes2017} must be acknowledged. Regarding \textit{internal validity}, participant selection via professional networking may introduce selection bias, though standardised demonstrations and independent tool access were used to mitigate evaluator influence. \textit{External validity} is constrained by the sample size ($n=14$) and focus on a single case study; results may not fully generalise to junior analysts or specialised sectors like healthcare. Furthermore, reliance on a specific model (Mistral Medium 3.1) impacts reproducibility, as results may fluctuate with model updates. In terms of \textit{construct validity}, our evaluation focused on perceived utility and trust rather than a direct empirical measure of ASTRAL's architectural reconstruction accuracy or completeness. Similarly, while high consistency in Likert ratings (4.00–4.64) suggests \textit{conclusion validity}, the small sample size limits the statistical power for detecting subtle nuances, and responses may be subject to central tendency bias.

To address these limitations, future work will involve a comparative evaluation of Phases 1 and 2 against industry-standard tools like the Microsoft Threat Modelling Tool \cite{MicrosoftTMT2024} and OWASP Threat Dragon \cite{OWASP2024ThreatDragon}, following the methodology in \cite{Granata2024}. This transition toward objective metrics for threat completeness and reconstruction accuracy will move the framework beyond practitioner perception and toward a more rigorous, domain-agnostic validation of the ASTRAL approach.

\section{Conclusion}
\label{sec:Conclusion}

Our study explored the application of multimodal LLMs in the context of architecture-centric security assessment for CPSs. Through the development and evaluation of the ASTRAL prototype, our research demonstrated that multimodal LLMs possess significant potential to automate and enhance existing CPS security assessment workflows. By leveraging both textual and visual system artefacts, ASTRAL markedly streamlines the identification of architectural structures, trust boundaries, and attack vectors. Crucially, our approach addresses the pervasive challenge of architectural incompleteness by reconstructing structured models from fragmented documentation, providing a robust foundation for downstream threat and risk evaluations. To support transparency and future collaborative research, the tool's source code has been made publicly accessible \cite{ASTRAL2026}.

An ablation study across three case studies validated ASTRAL's design. A systematic survey of fourteen security practitioners, including six CISOs, validated ASTRAL’s usefulness, trustworthiness, and reliability. 
High ratings for operational value, coupled with favourable feedback regarding improved workflow efficiency, enhanced analytical rigour, interface design, risk quantification, and decision support features, affirm ASTRAL's practical relevance.

Despite these positive outcomes, the study surfaced important directions for future work. Key recommendations from participants include improving transparency in threat modelling and risk scoring, establishing more comprehensive feedback mechanisms, and enabling richer integration with operational datasets and threat intelligence frameworks. Addressing these points will be instrumental in maximising practitioner confidence and extending the capabilities of LLM-enabled security assessment platforms beyond static documentation.

In summary, our study positions architecture-centric security assessment using multimodal LLMs as a promising and scalable foundation for informed cybersecurity decision-making. By bridging the gap between incomplete architecture and rigorous probabilistic analysis, ASTRAL provides a pathway toward more resilient and adaptive security postures in complex CPS environments.

\bibliographystyle{ACM-Reference-Format}
\bibliography{references}


\end{document}